\documentclass[10pt,conference,letterpaper]{IEEEtran}

\usepackage{times,amsmath,epsfig}
\usepackage{algorithmic,endnotes}
\usepackage[linesnumbered, ruled, vlined]{algorithm2e}
\usepackage{balance}

\usepackage{textcomp}

\usepackage[dvipdfm,CJKbookmarks=true,bookmarksopen=true,colorlinks,citecolor=black,linkcolor=black,anchorcolor=black,urlcolor=black] {hyperref}

\DeclareMathSizes{10}{8}{6}{5}

\title{Latency Bounding by Trading off Consistency in NoSQL Store: A Staging and Stepwise Approach}

\author{%
{Yuqing Zhu{\small $~^{\#1}$}, Philip S. Yu{\small $~^{*2}$}, Jianmin Wang{\small $~^{\#3}$} }%
\vspace{1.6mm}\\
\fontsize{10}{10}\selectfont\itshape
$~^{\#}$School of Software, Tsinghua University, Beijing 10084, China\\
\fontsize{9}{9}\selectfont\ttfamily\upshape
$~^{1}$zhu-yq@mails.thu.edu.cn\\
$~^{3}$jimwang@tsinghua.edu.cn%
\vspace{1.2mm}\\
\fontsize{10}{10}\selectfont\rmfamily\itshape
$~^{*}$Department of Computer Science, University of Illinois at Chicago, Chicago IL 60607, USA\\
\fontsize{9}{9}\selectfont\ttfamily\upshape
$~^{2}$psyu@uic.edu
}

\begin{document}
\maketitle

%\tableofcontents

\begin{abstract}
Latency is a key service factor for user satisfaction. Consistency is in a trade-off relation with operation latency in the distributed and replicated scenario. Existing NoSQL stores guarantee either strong or weak consistencies but none provides the best consistency based on the response latency. In this paper, we introduce dConssandra, a NoSQL store enabling users to specify latency bounds for data access operations. dConssandra \emph{d}ynamically bounds data access latency by trading off replica \emph{con}sistency. dConssandra is based on Cassandra. In comparison to Cassandra's implementation, dConssandra has a staged replication strategy enabling synchronous or asynchronous replication on demand. The main idea to bound latency by trading off consistency is to decompose the replication process into minute steps and bound latency by executing only a subset of these steps. dConssandra also implements a different in-memory storage architecture to support the above features. Experimental results for dConssandra over an actual cluster demonstrate that (1) the actual response latency is bounded by the given latency constraint; (2) greater write latency bounds lead to a lower latency in reading the latest value; and, (3) greater read latency bounds lead to the return of more recently written values.
\end{abstract}


\section{Introduction}%
The occurrence of NoSQL stores and cloud storage attracts attentions from both industry and academia, as the inevitable arrival of Big Data. NoSQL stores have been widely employed in a wide range of online services over the past few years. Web search, social networking and recommendation, stock trading, webstore, and gaming represent a few prominent examples of such services.

While very different in functionality, these services share three common underlying themes. First is their request for serving users in a timely fashion. Availability and latency is the most prominent factors for user satisfaction \cite{jHamilton}. User requests are to be satisfied within a specified latency target; or users leave. Latency causes penalties \cite{latencyCost, latencyCostSales, perfScale}. Second, scalability is the required properties as the big data challenge arrives. Scalability is also the preferable property of NoSQL stores over traditional database. The third is the geographical distribution of these services. Most of these services require a deployment up to multiple data centers. This geographical distribution lends data reliability and availability to these services.

In response to the three themes, NoSQL stores are built with availability, scalability and reliability properties. Replication is one key technique to achieve these favorable properties. According to the CAP theorem \cite{brewer:cap}, NoSQL stores have to sacrifice consistency for availability. Thus, many NoSQL stores choose to guarantee only eventual consistency for online service, e.g., Dynamo \cite{dynamo} and Voldemort \cite{voldemort}.  While modern databases emphasize correctness, completeness and thus consistency, NoSQL stores challenge modern databases by abandoning consistency for availability.

However, eventual consistency cannot satisfy all application scenarios. The unknown consistency status on eventual consistency also greatly complicates application development. Thus some researchers pointed out that, it is not fair that NoSQL stores guarantee only eventual consistency \cite{brewerCap, raghuCap}, when the database community has abundant techniques ready for reference \cite{stonebrakerBlog}.

On the one hand, strong consistency must be guaranteed for developers when required. Some NoSQL stores guarantee strong consistency \cite{bigtable, hbase}, but they have to sacrifice availability under the network-partition circumstance, which is a highly costly choice. On the other hand, requests must be responded within a given latency tolerable to users \cite{kersten:deluge}. Some recent work makes effort in reducing data access latency and guaranteeing strong consistency at the same time \cite{ibmPaxos}. But the response latency is inevitably impaired for bringing replicas consistent, as compared to that guaranteeing only weak consistency. As consistency is in a trade-off relation with availability and latency \cite{tradeoffExp, abadiBlog}, the system should maximize consistency as much as possible within this given latency. Hereafter, a strong consistency request can be responded in a longer latency.

In this paper, we present \emph{dConssandra} that supports latency-bounded operations with the best possible consistency. A better consistency status indicates a shorter latency for the consistent read; or, a better consistency status means a more recent value read within the same latency. The maximum latency tolerable for a service is easily known, and applications can tolerate varied degrees of replica consistency and varied magnitudes of latency \cite{raghu:keynote}. We thus propose a data access API with latency bound specification. Within the specified latency, it is guaranteed that a response be returned and consistency be maximized. Our key contributions in dConssandra include:%\vspace{-5pt}
\begin{itemize}
  \item a replication strategy that has multiple stages to enable synchronous and asynchronous replication on demand, and improve consistency as needed,%\vspace{-5pt}
  \item a storage architecture that allows flexible execution of the staged replication process, and supports operations with latency bounds,%\vspace{-5pt}
  \item a scheme to decompose the staged replication process into minute steps, so that the read/write execution latency can be measured and approximated by these minute steps, and%\vspace{-5pt}
  \item an algorithm dCON that computes a write subset and a step subset to be processed for the best consistency and within the given latency bound.%\vspace{-5pt}
\end{itemize}

We have implemented dConssandra as part of the widely-known open-source project Cassandra \cite{cassandra}. We deploy and evaluate dConssandra in a real cluster of nodes. Our results illustrate that (1) the actual response latency is bounded by the given latency bound; (2) greater write latency bounds lead to a lower latency of the consistent read (reading the latest value); and, (3) greater read latency bounds lead to the return of more recently written values.

In the rest of the paper, Section \ref{sec:systemOverview} provides a high-level overview of the dConssandra system. Section \ref{sec:tradeoffInDcon} describes the dConssandra replication strategy, and analyzes the consistency versus latency trade-off. Section \ref{sec:storageStructures} demonstrates dConssandra's storage structure. The decomposition scheme and the dCON algorithm are presented in Section \ref{sec:latencyConstrain}. Section \ref{sec:experiment} evaluates dConssandra implementation over a PC cluster and demonstrates the experimental results. Related work is summarized in Section \ref{sec:related}. We conclude and discuss future works in Section \ref{sec:conclude}. \vspace{-5pt}
\section{Overview}%
\label{sec:systemOverview}%
Prior to illustrating the operations with latency-bound specifications (\textsection \ref{sec:api}), we first briefly summarize dConssandra's data model (\textsection \ref{sec:dataMdl}). Data model is the basis of replication strategy and storage architecture design. Application developers can operate on the data model through dConssandra's API. Each operation is guaranteed to return within the given latency specification. At the end of this section (\textsection \ref{sec:systemArch}), we overview the system architecture of dConssandra.\vspace{-5pt}
\subsection{Data Model}%
\label{sec:dataMdl}%
We assume the flexible table model similar to that of Cassandra \cite{cassandraPaper}, BigTable \cite{bigtable}, and HBase \cite{hbase}. Data are organized into \emph{tables} as in BigTable, or \emph{keyspaces} as in Cassandra. A table consists of a set of \emph{rows}, which are uniquely identified by row keys. A table is defined with a set of \emph{column families}. Data in each row is organized into column families. A row may have no data in some column families. Row data in a column family is further organized into \emph{columns}, each of which is a name-value pair. Some row may have more columns in a column family, while some may have fewer or none.

Tables are partitioned into \emph{tablets} according to row keys. Tablets are replicated. Each tablet replica is separately distributed to a node. On a node, rows of a tablet are organized and stored in the unit of column family, which is called \emph{SSTable}. In the following, we refer to an SSTable as a data object, which is the most basic unit for replication and storage.  Note that column is the basic unit for operation.
\subsection{Operations  with Latency Bound Specification}%
\label{sec:api}%
The key API mainly supports three read/write operations, i.e. \textbf{\emph{read}}, \textbf{\emph{write}}, and \textbf{\emph{readTestWrite}}. The \emph{readTestWrite} operation cannot be easily implemented in NoSQL stores guaranteeing eventual consistency, but it is naturally supported by dConssandra.

The basic data unit for these operations is column, which is referenced by \emph{table},\emph{ row key}, \emph{column family}, and \emph{column}. Here the column may also contain a super column name (as in Cassandra), if the column family is declared to be one with super columns. Besides the reference to the data unit, an operation must also be specified with the expected latency bound \emph{tBound}. The operation will return a response within the latency \emph{tBound}.

The \emph{write} and \emph{readTestWrite} operations can be specified with an \emph{ordered} parameter. If \emph{ordered} is evaluated to \emph{true}, the operation is an in-order operation. In-order operations are executed in the same arrival order with other in-order operations. If \emph{ordered} is false or not given, the operation has a lower processing priority than in-order operations, so its execution order is not guaranteed. The \emph{readTestWrite} operation must also be specified with two data units and three values. If reading the first data unit returns a value equal to the first given value, the second data unit is set to the second value; otherwise, the second data unit is set to the third value. Note that, the two data units must reside in the same row and column family, i.e., the same tablet. In the following, we also call a \emph{readTestWrite} as a write.

\textbf{Write Bound.} Writes always get accepted and responded in dConssandra. If a write is specified with an infinite response latency, all previous ordered writes on the same data unit are guaranteed to be applied before this new write, leading to a following instantaneous read returning a recent value. The minimum processing latency for a write is the time dConssandra can guarantee the durability of the write, e.g., recording the write in a node-local log or forwarding the write to an adequate number of nodes. A response latency smaller than the minimum processing latency would be ignored. A write with an intermediate response latency will lead to partial processing of this write and/or previously unprocessed writes. A larger response latency leads to a more recent value for a following instantaneous read, or a shorter processing latency for a following consistent read with an infinite latency specification.

\textbf{Read Bound.} A read is guaranteed to be forwarded to its corresponding replica in dConssandra. If a read is specified with an infinite response latency, all previous ordered writes on the same data unit are guaranteed to be applied before this new read. That is, the read returns the most recent value by ordered writes. The minimum processing latency for a read is the time that the receiving replica processes the read locally. If there are previous unexecuted writes on the corresponding data unit, dConssandra returns NULL for the read, if given a response latency smaller than or equal to the minimum processing latency; otherwise, dConssandra returns what the replica reads locally. A read with an intermediate response latency will lead to partial processing of previously unexecuted writes before processing the read. A larger response latency leads to a more recent value returned by the read.
\begin{figure}[t]
      \centering
      \includegraphics[width=0.5\textwidth]{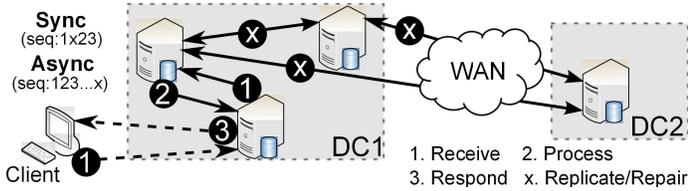}
      \caption{Sync/async replication in a multi-DC environment.}\vspace{-15pt}
      \label{fig:existingReplicationStrategies} %% label for entire figure
\end{figure}%

\textbf{Use Cases.} Consider the following use cases for the above operations. Studies report that web users leave a website if the web page does not display in three seconds. After subtracting all the time required for foreground processing, the time left for background data access becomes less than 100 milliseconds. Due to importance of the time limit, weaker consistency and stale data become tolerable. With the above latency-bounded operations, developers can specify a latency bound with 100 milliseconds for the relating data access operation. This will guarantee the page displays before users get impatient and leave. After the web page gets displayed, the user becomes less sensitive to the latency and the contents become important. For example, a user may want to read the most recent tweets from a few closely following friends later. In this case, consistency becomes important so that the most recent tweets are actually retrieved. With the above operations, developers can specify an infinite latency bound to get a consistent value.%\vspace{-6pt}
\subsection{System Overview}%
\begin{figure}[t]
      \centering
      \includegraphics[width=0.5\textwidth]{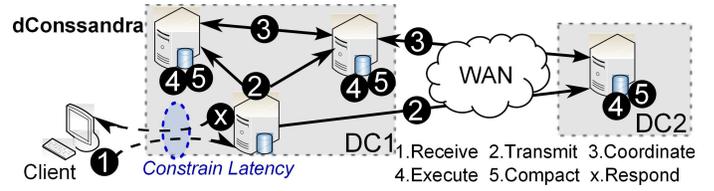}
      \caption{dConssandra's staged replication in a multi-DC environment.}\vspace{-15pt}
      \label{fig:dconReplicationStrategy} %
\end{figure}

\label{sec:systemArch}%
dConssandra consists of connected nodes with computing and storage resources. Nodes are distributed over one or more data centers. The system architecture is \emph{symmetric}. Nodes and replicas can be equally accessed. If an operation request arrives at a receiving node without the corresponding data, this request is forwarded to the closest node holding the replica. Operation results are forwarded back to the receiving nodes and then returned to the requester. The operation latency is computed from when the operation arrives at the receiving node till when the receiving node responds to the requester.

dConssandra's replication strategy breaks the replication process into six major stages (Figure \ref{fig:dconReplicationStrategy}). The six stages include \emph{reception}, \emph{transmission}, \emph{coordination}, \emph{execution}, \emph{compaction} and \emph{acquisition}. The stages are explained in the next paragraph. Stage is the basic unit to guarantee durability and failure tolerance. In comparison, among existing replication strategies (Figure \ref{fig:existingReplicationStrategies}), synchronous schemes replicate data (stage \emph{x}) right after the reception (stage \emph{2}), while asynchronous schemes respond (stage \emph{3}) first and replicate/repair data when necessary.

In dConssandra, the processing of a consistent write goes through all the stages except acquisition before response. The processing of a consistent read can go through all the stages except reception. The reception stage confirms the receipt of a write, and the acquisition stage acquires the data value to be read. The processing of any write always contains the reception stage, while that of any read contains the acquisition stage. The response latency is bounded by processing the maximum number of stages possible within the time bound. The \emph{x} stage \emph{Respond} may follow any other stage to end a processing procedure. We will detail how dConssandra replication strategy enables the flexible trade-off between consistency and latency in \textsection \ref{sec:tradeoffInDcon}. The dConssandra storage architecture and the processing flow, which together guarantee durability and availability regardless of whichever stage subset is selected, will be presented in \textsection \ref{sec:storageStructures}.

In order to bound latency, the processing time needs to be predicted. dConssandra borrows the idea from Riemann integral in latency prediction. It decomposes stages further into minute steps. The processing time of each step is approximated and predicted by a linear function. dConssandra estimates the latency by summing up the approximated processing time of all involving steps. Statistics are collected on the processing of each step to enable approximation. The system has warm-up time before estimation becomes effective. In addition, we need to choose the steps and stages carefully so that the system can guarantee the best consistency within the time limit. In \textsection \ref{sec:latencyConstrain}, we will elaborate on how the stages are decomposed into steps, so that latency prediction and bounding become possible. We also present the dCON algorithm to effectively choose the right set of steps for within-time processing. This choice is made when the NoSQL operation is received and before any processing is done (as indicated in Figure \ref{fig:dconReplicationStrategy}).
\section{Staged Replication Strategy \& the Tradeoff}%
\label{sec:tradeoffInDcon}%
Replication is an important technique to guarantee availability, scalability and reliability in NoSQL store. Replication strategy defines how the system guarantees the consistency between replicas. The process of bringing replicas consistent inevitably causes extra latency over that of single copy update execution. This extra latency becomes so prominent in the large-scale system that applications would rather sacrifice consistency for shorter latency. In this section, we first elaborate dConssandra's replication strategy (\textsection \ref{sec:dconReplicationStrategy}). We then discuss how replica consistency can be flexibly traded off for latency in dConssandra replication strategy (\textsection \ref{sec:consistencyVsLatencyTradeoff}).\vspace{-5pt}
\subsection{Staged Replication Strategy}%
\begin{figure}[t]
     \centering
      \includegraphics[width=0.49\textwidth]{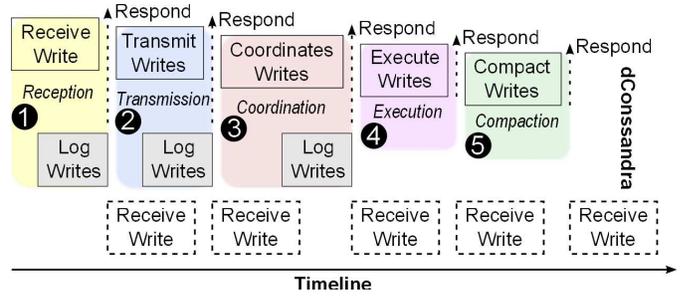}
      \caption{dConssandra's staged replication process for write.}\vspace{-16pt}
      \label{fig:dconReplicationProcess} %
\end{figure}

\label{sec:dconReplicationStrategy}%
dConssandra replication strategy allows \emph{update-anywhere}, \emph{eager synchronous} \textbf{and} \emph{lazy asynchronous} replication simultaneously. Figure \ref{fig:dconReplicationProcess} demonstrates the dConssandra replica control process for a write. The processing flow of a write can contain at most five stages, though it must always contain the reception stage. A consistent write with eager replication goes through all the five processing stages. Taking fewer stages leads to a less consistent write with lazy asynchronous replication.
\begin{figure}[t]
      \centering
      \includegraphics[width=0.49\textwidth]{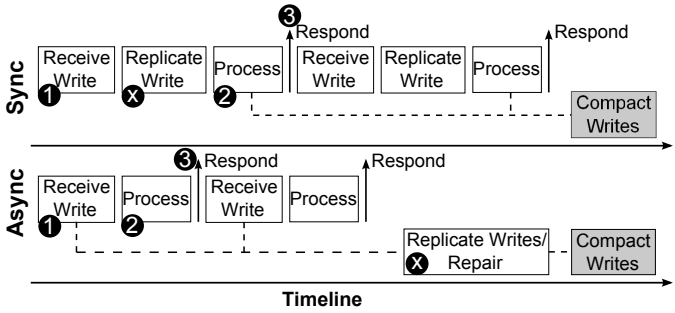}
      \caption{Existing replication processes for NoSQL stores.}\vspace{-16pt}
      \label{fig:existingReplicationProcesses} %% label for entire figure
\end{figure}
Read request processing can also contain at most five stages. The five stages are transmission, coordination, execution, compaction and acquisition. The first four stages handle writes previously received but not yet processed. The acquisition stage is required for the read processing to return the demanded value. Though only the acquisition stage actually processes the read request for the result, the first four stages can lead to a more consistent result returned by the read.

In the \emph{reception} stage, the write is received by one of the replica node. This write is then transmitted to all other replicas in the transmission stage. Since a write can be submitted to any replica, writes must go through a \emph{coordination} stage before execution. In this way, all replicas are guaranteed to execute the same write sequence in the \emph{execution} stage, thus avoiding conflict resolution. The processing of a write can also go through the \emph{compaction} stage. The compaction stage is not directly related to replica consistency, but this stage helps to speed up the \emph{acquisition} stage for a later read. This is due to the NoSQL storage architecture.

dConssandra orders a chosen set of stages as in Figure \ref{fig:dconReplicationProcess}. Though the processing can actually take the stages in any order, the order in Figure \ref{fig:dconReplicationProcess} guarantees the output from the previous stage is taken as input by later stages. For example, the reception stage always precedes the other stages in the processing; and the coordination stage always follows the transmission stage. Notice that, a response can be returned to the requester at the end of any stage. Writes for transmission and coordination stages are logged before response. The logging is for the purpose of durability and failure tolerance. A new write can be processed whenever the preceding request is responded.\vspace{-3pt}
\subsection{Consistency versus Latency Trade-off}%
\label{sec:consistencyVsLatencyTradeoff}%
Applications passively experience the choice of consistency versus latency trade-off by using a system with existing replication strategies. Existing replication strategies can be divided into asynchronous and synchronous categories (Figure \ref{fig:existingReplicationProcesses}). Synchronous replication strategy guarantees strong consistency, but the synchronous replication process leads to a long latency, especially in the cross-datacenter case. The asynchronous replication strategy can respond quickly, but the inconsistent replica states can be easily exposed to the following reads. The main difference between the two existing schemes is whether the data replication stage happens before response. Usually, a storage system either chooses the synchronous replication or the asynchronous one. Systems like Cassandra takes the synchronous replication with its \emph{ALL} consistency level, and the asynchronous replication with its other consistency levels.

dConssandra enables applications to actively and dynamically choose the consistency versus latency trade-off. dConssandra decomposes the general replication stage into transmission, coordination, and execution stages, which can be further configured with processing writes. This decomposition enables control of replica consistency (or latency) by selection of stages and writes. Figure \ref{fig:dconReplicationProcess} demonstrates dConssandra’s staged replication process for write.

Each write/read processing contains a subset of the five stages in dConssandra. Besides, the number of writes to be processed can be decided individually for each stage. The longer latency is specified, the more stages and the more writes can be chosen; and vice versa. As the selection and processing of stages are in the order specified in Figure \ref{fig:dconReplicationProcess}, we can observe that the more stages means the more executable or executed writes. Executing more writes, or producing more executable writes, increases the probability that a following read returns a consistent value. The only way to increase this probability is to decrease the number of unexecuted writes, and shorten the time for keeping unexecuted writes for the most consistent replica. In dConssandra, this probability can be controlled directly by executing more stages and writes, or indirectly by latency bound.

A consistent read usually contains all five stages, unless no writes need to be processed in the first four stages. With the best consistency, a consistent read needs only to process the acquisition stage. The better the consistency status is, the fewer the unexecuted writes are left. In other words, the fewer stages and writes a consistent read needs to process, indicating a faster response. On the other hand, the better the consistency status is, the fewer writes are left unprocessed; thus, the more recent value can be returned by a read within the same latency bound.

Among applications requiring strong consistency, some can tolerate long writes, and some long reads. With dConssandra, the strong consistency can be guaranteed through either consistent write or consistent read, i.e., write/read with infinite latency bounds. Applications tolerating long writes can specify consistent writes and fast reads, while those tolerating long reads can specify fast writes and consistent reads. Applications not caring consistency can specify their maximum tolerable latency, so that future operations can have a stronger possibility to read a consistent values. Herein, application developers can actively choose the desirable latency vs consistency trade-off with dConssandra.\vspace{-6pt}
\section{Storage Structures Supporting Stages}%
\begin{figure}[t]
  \centering
  \includegraphics[width=0.49\textwidth]{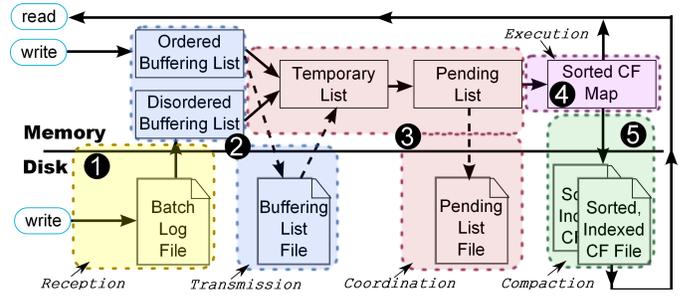}
  \caption{Storage structures and processing flow on one node.}\vspace{-15pt}
  \label{fig:storageArch} %
\end{figure}

\label{sec:storageStructures}%
In order to support the phased replication strategy, dConssandra implements a new storage architecture. As memory is becoming cheaply abundant and computing nodes are equipped with large memory, dConssandra exploits a good usage of memory and adopts the storage architecture demonstrated in Figure \ref{fig:storageArch}. With a symmetric system architecture, every node in dConssandra has the same storage structure. In the following, we first detail the design of dConssandra's storage structures in \textsection \ref{sec:storageStructureDetail}. How dConssandra preserves the favorable system properties of durability and failure tolerance is discussed in \textsection \ref{sec:durabilityNTolerance}.\vspace{-6pt}
\subsection{Storage Structures}%
\label{sec:storageStructureDetail}%
dConssandra allows the processing of read/write to stop at the end of any replication stage. For the purpose of durability and failure tolerance, dConssandra has storage structures to hold writes durably for each replication stage.

\emph{Batch Log File(bl-file)} for the \emph{reception} stage stores writes in disk. The order to execute writes in bl-file is not guaranteed. bl-file enables dConssandra to receive writes even under network partition, thus providing some form of availability. After network partition is fixed, writes in bl-file are sent to other replicas, following the write execution process in normal cases, except that these writes are not guaranteed with execution order. That is, the \emph{ordered} parameter is set to \emph{false} for writes received under network partition.

Without network partition, the received writes can be transmitted to all reachable replicas in the \emph{transmission} stage. Before transmission, writes requiring to be executed in order are stored in \emph{Ordered Buffering List (ob-list)}, while those not requiring order are in \emph{Disordered Buffering List (db-list)}. The \emph{ordered} parameter signifies whether a write is an ordered one or a disordered one. The ob-list and db-list constitute the \emph{Buffering List (b-list)}. Thresholds can be set for b-list. If the thresholds are reached, newly received ordered writes are logged in the \emph{Buffering List File (b-file)}, while disordered writes are appended to the bl-file. When there are not enough writes for coordination in b-list, writes in b-file are first retrieved and then those in bl-file.

The \emph{coordination} stage exploits \emph{Temporary List (t-list)} to temporarily store writes under coordination. Coordinated writes are stored in \emph{Pending List (p-list)}, pending for execution. If not all replicating nodes participate in the coordination stage, the coordinated writes are also logged in the \emph{Pending List File (p-file)}. Temporarily unavailable nodes can later request the file for a catch-up.

In the \emph{execution} stage, an executed write updates one or more columns of the corresponding row in a certain column family. The updated column values are stored in its \emph{Sorted Column Family Map (cfMap)}, which corresponds to an SSTable. Usually coordinated writes in p-list are executed immediately since the execution takes a relatively short time and directly improves consistency status.

Either when the \emph{compaction} stage is initiated or a cfMap reaches its size threshold, a cfMap is flushed out as a \emph{Sorted, Indexed Column Family File (cfFile)}. There may be more than two cfFiles for one column family simultaneously if a large number of writes pour in. The cfFiles can overlap in row keys. A cfFile compaction, similar to the leveled compaction of Cassandra, is initiated when the number of cfFiles reaches the configured limit, or when a compaction stage is initiated actively. Actively initiating compaction at spare time can effectively avoid the sudden surge of resource consumption and the huge performance degradation \cite{ycsbplus} due to the passive large-scale compaction.

In the \emph{acquisition} stage, read requests are first directed to the cfMap, and then the cfFiles. The cfMap is sorted according to row key, and the cfFiles are indexed to fasten the reading process.

The multi-level in-memory structures enable staged and batch processing over costly processes such as network communication, coordination, read-test-write execution, and compaction. Batch processing can effectively reduce unit processing cost by amortization. As only cfMap and cfFile are readable, efforts must be made towards executing more writes for cfMap and cfFile storage. Though compacting cfMaps does not directly increase consistency, it reduces the latency to read.

The dConssandra replication process can thus be summarized as follows. A replica may just put a newly received write in the b-list and send it if given a tightly bounded response latency. If the latency limit allows for more operations, the replica can execute coordinated writes in p-list, compact the executed writes, request the leader for a write sequence coordination process, or transmit disordered writes in bl-file. More operations and steps can be taken on demand as the latency limit allows for stronger consistency guarantee. The read processing follows a similar process, except that it does not have the reception stage and that it must have the acquisition stage. A replica with spare resources, e.g., CPU cycles, network bandwidth and memory, can similarly take some or all of the stages to actively improve its consistency status for future requests.
\begin{table}[t]
\centering
\caption{Read/Write Execution Steps}%
\label{tbl:steps}%
\small
\begin{tabular}{|c|l|c|c|} \hline
\textbf{$\#$} &\multicolumn{1}{c|}{\textbf{Step Description}}&\textbf{Cost}&\textbf{Stage}\\ \hline
1 &flush $i_1$ writes to bl-file & $f_d(i_1)$  & Reception\\ \hline
2 &append $i_2$ writes to b-list & $f_a(i_2)$ & \\
3 &read $i_3$ writes from bl-file & $f_r(i_3)$ & \\
4 &transmit $i_4$ writes &  $f_t(i_4)$ & Transmission \\
5 &append $i_4$ writes to b-list & $f_a(i_4)$ & \\
6 &flush $i_5$ writes to b-file & $f_d(i_5)$ & \\
7 &transmit ack &  $f_t(1)$ & \\ \hline
8 &read $i_6$ writes from b-file & $f_r(i_6)$ & \\
9 &transmit $i_7$ writes from b-list&  $f_t(i_7)$ & \\
10 &append $i_7$ writes to t-list & $f_a(i_7)$ & \\
11 &transmit $i_8$ writes &  $f_t(i_8)$ & \\
12 &align $i_9$ writes &  $f_c(i_9)$ & Coordination\\
13 &transmit $i_{10}$ writes &  $f_t(i_{10})$ & \\
14 &insert $i_{11}$ writes into p-list & $f_i(i_{11})$ & \\
15 &flush $i_{11}$ writes to p-file & $f_d(i_{11})$ & \\
16 &transmit ack &  $f_t(1)$ & \\ \hline
17 &execute $i_{12}$ writes &  $f_e(i_{12})$ & Execution \\ \hline
18 &compact $j_1+k_1$ rows &  $f_C(j_1,k_1)$ & Compaction \\ \hline
19 &read $j_2+k_2$ rows & $f_R(j_2,k_2)$ & Acquisition \\ \hline
\end{tabular}\vspace{-10pt}
\end{table}
\begin{table}[t]
\centering
\caption{Conditions for Read/Write Execution Steps}%
\label{tbl:condt}
\small
\begin{tabular}{cl|cl} \hline
\textbf{Seq $\#$} &\multicolumn{1}{c|}{\textbf{Condition}} & \textbf{Seq $\#$} &\multicolumn{1}{c}{\textbf{Condition}} \\ \hline
\textcircled{a} & Is a write request & \textcircled{b} & Is a read request \\ \hline
\textcircled{c} & Is connected to & \textcircled{d} & Is some replica\\
 & a quorum of replicas &  & failed\\ \hline
\textcircled{e} & Has writes in b-list & \textcircled{f} & Has writes in p-list \\ \hline
\textcircled{g} & Has writes in bl-file & \textcircled{h} & Has writes in b-file \\ \hline
\textcircled{i} & Has space in b-list & \textcircled{j} & Has space in p-list\\ \hline
\textcircled{k} & \multicolumn{3}{l}{Has writes in cfMap or cfFile number exceeds limit} \\ \hline
\end{tabular}\vspace{-15pt}
\end{table}
\subsection{Durability and Failure Tolerance}%
\label{sec:durabilityNTolerance}%
For durability and failure tolerance purpose, b-list writes must be sent to at least a quorum of replicas before response. The truth of this statement comes naturally from the quorum read-write rule.  We assume that no more that half of the the replicas should simultaneously fail. Thus the number of replicas is important to this purpose. Given a node's failure probability is $p_i$ and a safe system is one that have only $p_f$ probability of failure, the reliable replica number must be greater than the $r$ that solves $p_f=\frac{p_i^{[r/2]}-p_i^{r+1}}{1-p_i}$. Under network partition, a b-list write cannot reach a quorum of other replicas. The receiving replica must store it in the bl-file. If the present bl-file contains only sent writes and there is no incoming write at the moment during the maintenance process, the bl-file is discarded and a new bl-file is created.

The coordination stage guarantees all replicas always execute a prefix of the same write sequence. The benefit here is to avoid the troublesome conflict resolution. Paxos-based protocols \cite{ibmPaxos} enable cohort-based write coordination, which avoids the bottleneck drawback and provides good failure-tolerance feature. Our coordination adopts the Paxos-based approach, but differentiates from the recent proposal \cite{ibmPaxos} in the following aspects: (1) no centralized coordination service is exploited for consideration of scalability; (2) the coordination leader changes on every commit and is decided dynamically to allow better load balancing and thus scalability; (3) multiple writes are involved in one coordination process, in comparison with the normal coordination-on-each-write approach. NTP (Network Time Protocol) is exploited to synchronize nodes' time. The leader node aligns writes from different replicas based on their timestamps. Note that NTP can only coordinate nodes' time to a certain precision, so the leader node randomly decides the sequence of writes with the same timestamp. If the chosen leader for a coordination process fails, the coordination falls back to the ordinary case of Paxos protocol.

During coordination, if there is any failed replica, the leader requires all participating replicas to store the coordinated write sequence in the p-file. A failed replica can catch up with other replicas during the coordination process. The p-file gets truncated once its size exceeds the threshold for a missing-write catch-up process. When the p-file grows too large, it is more efficient for a replica to recover and catch up by copying cfMap and cfFile than using p-file records. The p-file size threshold for truncation is the size of p-list.
\begin{figure}[t]
  \centering
  \includegraphics[width=0.39\textwidth]{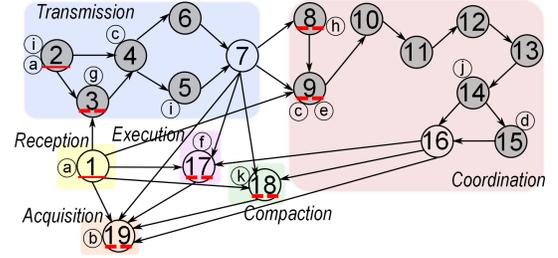}
  \caption{Graph of ordered execution steps.}\vspace{-15pt}
  \label{fig:paths} %% label for entire figure
\end{figure}
\section{Stepwise Bounding of Response Latency}
\label{sec:latencyConstrain}%
Our main idea to bound response latency is decomposition into and approximation by small steps. This is similar to the idea of Riemann integral. We break the process into minute steps, each of which relates to a limited range of processing and time. This small granularity makes latency prediction possible. The total latency is the summation of all step latencies. Our prediction of unit processing time for a step is based on the intuition that recent statistics have larger impacts on the following event, as workloads usually do not drop or soar suddenly (observing in the granule of seconds), even under sudden activity spikes.

dConssandra computes the step subset for execution after receiving the operation and before processing it. The computation is based on the latency estimation and the stage boundary. The stage boundary must be complied for the sake of durability and failure tolerance. The computation not only decides the step subset, but also the set of writes/reads to be processed by each step. In the rest of this section, we first introduce how we decompose stages into steps. We then analyze how consistency can be maximized in the process of latency bounding. Finally, we present the dCON algorithm that does the computation.
\subsection{Decomposing into Steps}%
\label{sec:step}%
With the above storage architecture, dConssandra's phased replication stages can be further decomposed into 19 steps as shown in Table \ref{tbl:steps}. The possible execution order of these steps and the necessary pre-conditions for step execution are demonstrated as the directed graph of Figure \ref{fig:paths}. Each step is represented as a node in the graph. A directed edge represents a feasible order of step execution. Rounded alphabets next to a step node are the required conditions. All conditions next to a step must be satisfied before executing the step. Descriptions of these conditions are listed in Table \ref{tbl:condt}. Nodes \emph{1} and \emph{2} are eligible starting points of a write execution process, while nodes with dashed underlines are eligible to start a read process. Only hollow nodes are legal ending points of an execution process for the sake of durability and failure tolerance, while filled nodes are only intermediate transition steps.

\textbf{Minimal processing.} Consider a write request given momentary response latency limit. The replica receiving the write will record it in its b-list, send it to a quorum of other replicas and respond, forming a path of (\emph{2,4,5,7}) or (\emph{2,4,6,7}) in the graph; or, on network partition, the replica records the request in the bl-file and returns, forming the path of (\emph{1}). A replica processing a read request with momentary response latency limit just returns the corresponding value in cfMap or otherwise in cfFiles, thus leading to a path of (\emph{19}).

\textbf{Maximal processing.} On receiving a write with infinite latency bound, the receiving replica can additionally send out writes in b-file and bl-file, request the leading replica for a write sequence coordination, execute all coordinated writes, and compact the cfMap with the cfFile. The path of (\emph{2,3,4,5,7,8,9,10,11,12,13,14,16,17,18}) represents the above processing procedures. A read with infinite latency bound can also lead to a long processing path. The receiving replica starts processing by requesting all other replicas to send out their bl-file and b-file writes, resulting in the path of (\emph{3,4,5,7,8,9,10,11,12,13,14,16,17,18,19}). This long path enables the return of the consistent value. The read/write request with an intermediate latency bound is correlated with a path that has a length between that of the maximal and the minimal.\vspace{-5pt}
\subsection{Consistency Maximization and Path Cost}%
\label{sec:optPath}%
According to the analysis in \textsection \ref{sec:consistencyVsLatencyTradeoff}, dConssandra must take as many stages for an operation processing as possible. Besides, the more writes are processed in each step, the better. So to bound response latency and maximize consistency requires finding in the graph the longest path which satisfies some conditions. Writes correlating with each node in the found path must also be decided.

In the graph of Figure \ref{fig:paths}, a path with appropriate starting and ending points is a legal path, given that its path cost $C_p$ does not exceed the given response latency bound $t_r$. Satisfying all node conditions is the underlying assumption for legal paths. Legal paths are candidates for solutions to the consistency maximization problem. The cost summation of all steps related with a path is referred as its \emph{path cost}. Each step (node) has a related cost function as shown in Table \ref{tbl:steps}. The execution of a step incurs costs that can be computed by the corresponding cost function. There are nine node cost functions, each of which can be modeled as a linear function with the form $f(x)=a+bx$. Given the linear property of node cost functions, the path cost $C_p$ thus takes a linear form:\vspace{-6pt}
\begin{equation}
C_p=I_p^TA_p+B_p^TX_p\vspace{-6pt}
\label{eq:pcost}
\end{equation}

The $i$th element of vector $A_p$ is the parameter $a$ for the $i$th node on the path. Similarly, the $i$th elements of vector $B_p$ and $X_p$ are the parameter $b$ and the corresponding $x$ for the $i$th node on the path. $I_p$ is an all-one vector with the same dimension as $A_p$. The function parameters $a$ and $b$ can be computed and decided through statistics collected during system runtime. Each node records the processing time and the corresponding $i$ for each step. In computing the cost function, we set higher weights for recent records and lower weights for earlier records. We discard the collected statistics after the computation. Note that each node has its own learned cost function weights. The path cost computation and the write number estimation always exploit the latest learned cost functions.

Given a response latency bound $t_r$, the latency for the write category must include the acknowledgement time, while that for the read category must include the result transmission time. Assuming the transmitted data size is $s$. For the write category, $s$ is equal to the size of an acknowledge message; for the read category, it is equal to the result set size. Thus we have:\vspace{-6pt}
\begin{equation}%
\label{eq:readwriteLimit}%
    t_r\geq f_t(s)+C_p\vspace{-6pt}
\end{equation}

On the other hand, producing more executable and executed writes requires more steps to be taken, i.e., a longer legal path. This is contradicting with the response latency limit $t_r$, and the limited resources, e.g., network communication latency and CPU cycles. To solve the consistency maximization problem given a latency bound, we need to find the longest legal path producing more executable writes and satisfying the path cost constraint.\vspace{-5pt}
\subsection{dCON Algorithm}%
\label{sec:algo}%
We can make the following analysis based on the graph in Figure \ref{fig:paths}. Among all hollow nodes that serve as ending points for legal paths, paths taking nodes $7$, $16$ and $17$ directly improve consistency and produce more executable writes. Whenever time allowing and conditions met, node $17$ must be taken first. Nodes $2,4,5,7$ are almost always taken on writes, while node $1$ should only be taken under network partition, as it causes extra cost. For read, node $19$ must be taken. The more writes involve in steps of the \emph{coordination path} starting at node $8$ and ending at node $16$, the smaller average coordination cost for writes it will lead to. Note that the write numbers $X_p$ in equation \eqref{eq:pcost} relate with each other as indicated by the above guidelines. Summarizing these guidelines for the best consistency and referring to Equation \eqref{eq:pcost} \& \eqref{eq:readwriteLimit}, we obtain the algorithm on finding the legal path and the write numbers for each node in the graph, as shown in algorithms \ref{algo:pdcon} and \ref{algo:dcon}. Note that readTestWrite is taken as an ordinary write except with a different execution cost estimation and function.

\textbf{Pre-dCON}. We assume every request arrives at the computing node with the corresponding replica. The replica decides which steps and their corresponding writes/reads can be processed within the given latency bound and producing the best consistency. The operation processing follows the decision. Given the category information and the latency bound, algorithm \ref{algo:pdcon} outputs a path representing a decision on steps. It does the pre- and post- processing for read and write respectively, according to equation \eqref{eq:readwriteLimit}, and then invokes dCON algorithm (algorithm \ref{algo:dcon}) to decide for more intermediate steps for execution. dCON algorithm is the main part to compute the decision.
\setlength{\textfloatsep}{0.2\baselineskip} %{0pt}
\begin{algorithm}[t]
\small
\KwIn{$t_r$, isWrite}%
\KwOut{A path of nodes $S_P$}%
Set $S_P\leftarrow {\O}$\;%
\eIf(\tcc*[f]{write category}){isWrite}{%
\eIf{not connected to a quorum of nodes}{
Set $i_2=1$\;%
Add node $1$ to $S_P$\;%
}{
Set $i_4=i_7=1$\;%
Add nodes $2,4,5,7$ to $S_P$\;%
Compute $S_P$'s path cost $C_P$\tcp*[f]{$f(1)$}\;%
$S_P$=dCON($t_r-C_P$, $S_P$)\;%
}%
}%IF
(\tcc*[f]{read category}){%read category
Set $j_2$, $k_2$\;%
Add node $19$ to $S_P$\;%
Compute $S_P$'s path cost $C_P$\tcp*[f]{$f(j_2+k_2)$}\;%
$S_P$=dCON($t_r-C_P$, $S_P$)\;%
}%ELSE
\caption{Pre-dCON Algorithm} %
\label{algo:pdcon}
\end{algorithm}
\begin{algorithm}[t]
\small
\KwIn{$t_P$, $S_P$}%
\KwOut{A path of nodes $S_P$}%
Initiate all $i\_$s to 0\;
\If(\tcp*[f]{Executable writes}){$\textcircled{f}$}{%
Compute $S_P$'s path cost $C_P$\;%
Estimate the number $m_e$ of executable writes within $t_P-C_P$\;%
\If{$m_e>0$}{%
Set $i_{12}\leftarrow$ MIN$(m_e,$p-list write number$)$\;%
Add nodes $17$ to $S_P$\;%
}%IF m_e
    }%IF condition f
\If(\tcp*[f]{Coordinatable writes}){$\textcircled{c}\wedge\textcircled{e}\wedge\textcircled{j}$}{%
Compute $S_P$'s path cost $C_P$\;%
Estimate the number $m_c$ of writes coordinatable and executable within $t_P-C_P$\;%
\If{$m_c>0$}{%
Set $i_7\leftarrow$ MIN$(\frac{1}{r}m_c,$b-list size+b-file write number$)$\;%
Set $i_{12}\leftarrow i_{12}+m_c$\;%
\lIf{$i_7>$b-list size}{Add node $8$ to $S_P$}\;%
Add nodes $9$-$14$,$16$ to $S_P$\;%
}%IF m_c>0
}%IF condition c,e,j
\If(\tcp*[f]{Transmittable writes}){$\textcircled{c}\wedge\textcircled{g}$}{%
Compute $S_P$'s path cost $C_P$\;%
Estimate the number $m_t$ of writes transmittable within $t_P-C_P$\;%
\If{$m_t>0$}{
Set $i_4=i_3\leftarrow$ MIN$(m_t,$bl-file write number$)$\;%
%Set $i_7\leftarrow i_7+m_t$\;%
%Set $i_{12}\leftarrow i_{12}+m_t$\;%
Add nodes $3$-$5$, $7$ to $S_P$\;%
}%IF m_t
}%IF condition c,g
\If(\tcp*[f]{Compactable writes}){$\textcircled{k}$}{%
Compute $S_P$'s path cost $C_P$\;%
Estimate the number $m_z$ of compactions within $t_P-C_P$\;%
\If{$m_z>0$}{
Set $j_1$, $k_1$ based on $m_z$\;%
Add nodes $18$ to $S_P$\;%
}%IF m_z
}%IF condition k

\caption{dCON Algorithm(finding the legal path with maximized consistency)}%
\label{algo:dcon}
\end{algorithm}

\textbf{dCON.} The main idea of dCON algorithm is to compute the maximum number of processable writes within the latency bound for each stoppable stage. These stages include the execution of coordinated writes, the coordination of transmitted writes, the transmission of received writes and the compaction of executed writes.. The computation starts from the most contributive stage \emph{execution} to the least \emph{compaction}. Here contribution is considered with regard to the consistency status improvement. So the algorithm computes the maximum numbers of executable, coordinatable, transmittable and compactable writes one after another. Since these stages are correlated, the result for the previously computation is adapted in later stage computations.

In algorithm \ref{algo:dcon}, conditions are checked at lines $2,8,16$ and $22$ for the involving subpath before each computation. It first tries to execute as many coordinated writes as allowed within the latency constaint to directly improve consistency. Thereupon, it tries to coordinate as many transmitted writes as possible. In the time cost computation, the execution time of these newly coordinated writes is included. In computing coordinatable writes, steps $8$-$16$ and thus $i_6$-$i_{12}$ are involved. The assumption behind lines $12$ and $13$ of algorithm \ref{algo:dcon} is $i_7=i_8$, $ri_7=i_9=i_{11}=i_{12}$, $(r-1)i_7=i_{10}$, and that steps $9$-$14$ and $16$ are involved in the cost estimation. After that, dCON algorithm tries to transmit as many writes in bl-file as allowed within the latency bound, while these newly transmitted writes must also be considered for coordination and execution. Finally, compaction is added to the procedure if (1) there is enough time left; and (2) cfMap has writes, or the number of cfFiles exceeds the threshold. Given a latency bound large enough, all writes in bl-file, b-list, b-file, and p-list will be executed and data files are compacted, leading to a completely consistent and easily read value.%\vspace{-6pt}
\section{Implementation and Evaluation}%
\label{sec:experiment}%
We evaluate the effectiveness of dConssandra for a variety of aspects, with a discussion (\textsection \ref{sec:discussion}) ending this section. In particular, we will answer the following questions:%\vspace{-5pt}
\begin{itemize}
  \item Can dConssandra effectively bound latency?  (\textsection \ref{sec:latencyBound}) %\vspace{-5pt}
  \item What is the effect of trading off consistency? What is the influence of cross-datacenter bandwidth? (\textsection \ref{sec:tradeoff})%\vspace{-5pt}
  \item What overheads does dCON algorithm impose over the processing? (\textsection \ref{sec:overhead})%\vspace{-5pt}
  \item What is the correct latency bound choice for different workloads? (\textsection \ref{sec:diffWorkloads})%\vspace{-6pt}
\end{itemize}
\subsection{Experimental Setup}
The experiments were run on a cluster of 20 nodes connected with 1Gb ethernet. Each node is equipped with four 3.00GHz dual-core Intel Xeon processors, 16GB RAM, and one 256GB SATA disk. We use the Debian 4.0 (64-bit) as the host operating system. Debian Kernel 2.6.24-6~etchnhalf.4 is used for both test server and client operating systems, with disks formatted using the \emph{xfs} filesystem. We implement the dConssandra based on Cassandra 1.0.5. We exploit and modify the YCSB++ \cite{ycsbplus} cloud benchmark for testing.

dConssandra is symmetrically deployed on a shared disk for 18 nodes. The 18 nodes are divided into two groups. Each group consists of 9 nodes. We configure each group to function as a datacenter. The network bandwidth between two datacenters can be configured by software. Within each datacenter, there are two racks having 4 nodes and 5 nodes respectively. The log and the data directories for the running Cassandra are configured to one dedicated SATA disk on each node. We disable the unnecessary Cassandra features like hinted handoff and read repair for the dConssandra implementation. Both row and key caching are also turned off. The remaining two nodes among the 20 nodes are reserved as test nodes.\vspace{-5pt}
\begin{figure}[t]
    \centering
    \includegraphics[width=0.36\textwidth]{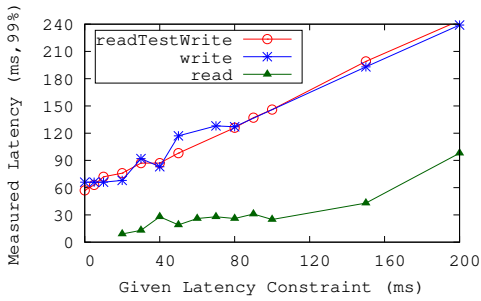}%\vspace{-10pt}
    \caption{Bounded latencies.}
    \label{fig:latencyBounds}
\end{figure}%
\begin{figure}[b]
  \centering
  \includegraphics[width=0.36\textwidth]{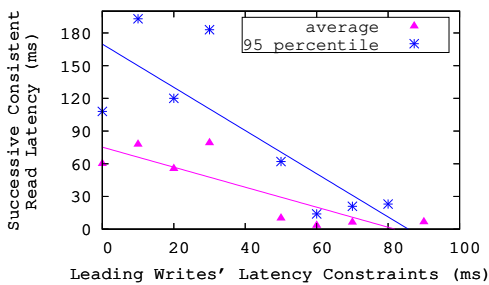}
  \caption{Consistent read following writes with varied latency bounds.}
  \label{fig:varwlongr} %
\end{figure}
\subsection{Bounded Response Latency}%
\label{sec:latencyBound}%
As for initialization, we run a set of 50 thousand inserts that are configured with unbounded latency bounds. The reason is three fold. Firstly, we must load the store with data to make the experiments fair. Secondly, the cost function can be trained in the process. Thirdly, the time needed to do a totally consistent insert must be estimated for later experiments.

In the first experiment set, we demonstrate how the latency bound is correlated with the measured latency. We vary the latency bounds for the API from 0ms to 200ms. The 0ms bound indicates that the operation must be responded as-soon-as-possible. Every latency bound for a write/read-test-write operation is an experiment, in which we run 10 thousand operations.

To test about read latencies, we also run a set of experiments that have read latency bounds varying from 0ms to 200ms. Different from the write experiments, in each experiment, we divide operations into 100 groups, each of which is 99 quick writes followed by a latency-bounded read. The writes and the read within the same group have the same key. That is, the read is requesting values just written. The reason behind this is to test whether read latencies can be bounded when further efforts for a better consistency can be made.

Figure \ref{fig:latencyBounds} demonstrates the measured latencies for the 99 percentile versus the given latency bounds. The measured latency for the 99 percentile means that 99\% operations have a latency below the measured one. From the figure, we can see that the measured latencies increase almost linearly as the latency bounds increase. The curves of these 99th percentile latencies indicate that most writes/read-test-writes/read have bounded latencies. We can also observe that the curves of the measured latencies do not completely coincide with the expected line. As the sizes of read values may vary, the estimation of reading costs has an even larger discrepancy with the actual measurement. This is due to the inaccuracy of the cost functions and the corresponding estimations. However, we note that refining the cost function is not the focus of this paper, where the objective is to show that the proposed dConssandra mechanism can effectively bound latency and trade off latency for consistency. Still Figure \ref{fig:latencyBounds} shows that\textbf{ the proposed dConssandra can control the general trend on achieving the latency versus consistency trade-offs}.\vspace{-5pt}
\begin{figure}[b]
  \centering
  \includegraphics[width=0.36\textwidth]{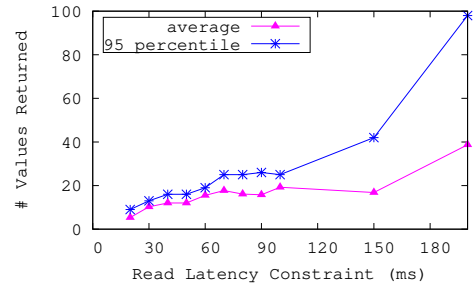}
  \caption{Number of Values returned by reads with varied latency bounds.}
  \label{fig:imwvarr} %% label for entire figure
\end{figure}
\subsection{Consistency versus Latency Trade-off}%
\label{sec:tradeoff}%
Having established the effectiveness of dConssandra at bounding operation latency, we now turn our attention to the correlation between latency and consistency. We first demonstrate that \emph{larger write latency bounds lead to lower latency for consistent reads}.  \textbf{Consistency is represented as the latency of a consistent read. The smaller latency, the stronger consistency.} We run a set of experiments, each of which has 100 groups of operations. Each group consists of 99 writes followed by a read with an infinite latency bound. That is, the read must return the latest value, which is also called Read-Your-Write consistency. The 99 writes for all groups in the same experiment are requested with the same latency bounds. We vary the write latency bound for each experiment from 0ms to 100ms. In Figure \ref{fig:varwlongr}, we draw a graph for the measured average and the 95th percentile latencies for reads, in correlation with the latency bounds of their preceding writes. From Figure \ref{fig:varwlongr}, we can observe that the larger the preceding write latency bound is, the shorter the measured read latency is. Although there are some variations, \textbf{the trend of the measured read latency is decreasing as larger write latency bounds are given}.

To check the consistency with regard to latency, we again run a set of experiments, each of which contains 100 groups of operations. Each group consists of 98 immediate writes followed by one read with a given latency bound. After the read, we initiate a consistent read so that later experimental groups are not affected by previous groups. The writes and the read in the same group access the same key. That is, we are trying to read the writes. For each set, the read latency bounds for all groups are the same. We vary the read latency bounds from 0ms to 200ms for each experiment. We are expecting that \emph{the larger the read latency bound is, the more values the read returns}. \textbf{Consistency is represented as the number values returned by an instantaneous read. The more values returned, the stronger consistency.} Figure \ref{fig:imwvarr} demonstrates the result. In the figure, the number of values returned by reads is positively correlated with the requested read latency bound. The curves start at 20ms. The reason is that no values are returned with a latency bound smaller than 20ms. On the whole, the expectation is validated. That is, \textbf{a larger latency bound for reads lead to a larger number of successful instantaneous reads}.

\begin{figure}[t]
  \centering
  \includegraphics[width=0.36\textwidth]{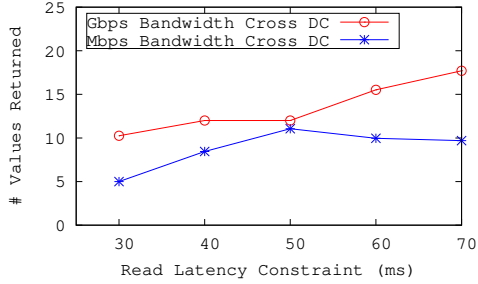}
  \caption{Consistency and latency under different cross-DC bandwidths.}
  \label{fig:xdc} %
\end{figure}

\textbf{Cross-Datacenter Bandwidth Influence.} In actual scenarios, the cross-datacenter communication is the major cause of latency when stronger consistency is required. The actual cross-datacenter bandwidth is Gbps. We delay all data transmissions to form a Mbps cross-datacenter bandwidth. Running the same set of experiments as in Figure \ref{fig:imwvarr}, we obtain the result as in Figure \ref{fig:xdc}. With Gbps cross-datacenter bandwidth, values are returned starting from the 20ms read latency bound. But the Mbps cross-datacenter bandwidth leads to a starting bound of 30ms. That is, achievable consistency in a higher cross-datacenter bandwidth is no longer achievable under a lower one. Figure \ref{fig:xdc} also shows that a higher cross-datacenter bandwidth indicates a stronger consistency within a given latency bound, as implied by the larger number of returned values for the Gbps cross-datacenter bandwidth.%\vspace{-5pt}
\subsection{Overhead of dCON Algorithm}%
\begin{figure}[t]
  \centering
  \includegraphics[width=0.36\textwidth]{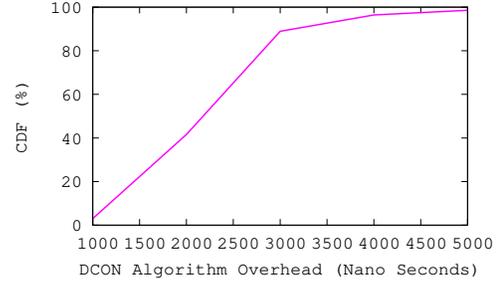}
  \caption{dCON overhead.}
  \label{fig:dconCost} %
\end{figure}

\label{sec:overhead}%
The user-specified latency usually is in milliseconds. The execution of dCON algorithm is in the path of operation processing. Therefore, the execution overhead of dCON algorithm must be small enough to be neglected. We log the execution time of dCON algorithm in the whole experiment.  Figure \ref{fig:dconCost} shows a CDF of dCON algorithm overhead to 99\%. As demonstrated, 99\% dCON executions are less than half a microsecond. As latency specification is in millisecond, this overhead is less than 5\textperthousand. The CDF curve has a long tail. The maximum overhead as logged is 0.2 millisecond. There is only one such sample. 99.7\% dCON executions are less than 1\% millisecond. This result shows that the overhead of dCON algorithm can almost be neglected. The reason that dCON has little overhead lies in its exploitation of local statistics and linear functions to approximate step processing time.%\vspace{8pt}
\subsection{Workloads with Varied Read/Write Ratio}%
\label{sec:diffWorkloads}%
Some applications care about single operation latency, while other care more about throughput. In this section, we vary the request workloads by choosing different read/write ratios. We assume the strong consistency is required. That is, reads must return the most recent value. We carry out two groups of experiments, each of which contains 20 thousand requests. The first group makes immediate write and consistent read requests, while the second makes consistent write and immediate read requests. Both groups guarantee consistent read values. Here, \emph{immediate} indicates a latency bound of \emph{zero}, and \emph{consistent} indicates a latency bound of \emph{unbounded}. The statistics for the two experiment groups are presented in Table \ref{tbl:morewrites} and Table \ref{tbl:morereads} respectively.
\begin{table*}[!tb]
  \begin{minipage}{0.5\textwidth}
\centering
\caption{Write Immediate Read Consistent}%\vspace{10pt}%
\label{tbl:morewrites}%
\small
\begin{tabular}{|c|c|c|c|c|}
  \hline
    W/R & Key & W-cost & R-cost & Throughput\\
    Ratio & Distribution & Ratio & Ratio & (ops/sec)\\ \hline
    9:1 & uniform & 95.55$\%$ & 4.45$\%$ & 92\\ \hline
    9:1 & zipfian & 97.49$\%$ & 2.51$\%$ & 86\\ \hline
    3:7 & uniform & 39.27$\%$ & 60.73$\%$ & 172\\ \hline
    3:7 & zipfian & 77.60$\%$ & 22.40$\%$ & 182\\ \hline
\end{tabular}
  \end{minipage}\vspace{-6pt}%
  \begin{minipage}{0.5\textwidth}
\centering
\caption{Write Consistent Read Immediate}%\vspace{5pt}%
\label{tbl:morereads}%
\small
\begin{tabular}{|c|c|c|c|c|}
  \hline
    W/R  & Key Avg.& W Avg.& R Avg.& Throughput\\
    Ratio & Distribution & Latency(ms) &  Latency(ms) & (ops/sec)\\ \hline
    1:9 & uniform & 2547.17 & 1.33 & 4 \\ \hline
    1:9 & zipfian & 2461.55 & 1.49 & 4 \\ \hline
    7:3 & uniform & 2561.24  & 1.82 & 1 \\ \hline
    7:3 & zipfian & 1907.86  &  4.78 & 1 \\ \hline
\end{tabular}
  \end{minipage}\vspace{-6pt}
\end{table*}

In the first group, we generate write/read requests in the ratios of 9:1 and 3:7. The key for each request is generated following the uniform and the zipfian distributions, respectively. Each row in Table \ref{tbl:morewrites} stands for a round, which consists of 20k requests. From the first two rows, we observe that the time cost percentages of write are quite close to the fraction of write requests (which is 90\%). Though consistent reads may lead to a long latency, immediate writes must be transmitted to and acknowledged by at least a quorum of nodes holding replicas, which may not be as immediate as indicated by \emph{immediate}. The 99th percentile latencies for read/write at either row are 41ms/914ms and 57ms/998ms, respectively. This leads to the indication that (1) batch coordination helps reduce average processing latency, and (2) write immediate and read consistent help reduce the total time cost if write requests are the majority in the workload.

The latter two rows in Table \ref{tbl:morewrites} have larger throughputs than the first two. This is because each round has the same total number of requests, and the total processing time increases with the number of writes. The write cost ratios for the latter two rows are different due to the distribution of requested keys. With zipfian distribution, some keys are accessed much more than the others, and the repeatedly accessed keys will reside in memory, thus leading to a shorter read time. Furthermore, as more read requests are initiated than write requests in the latter two rounds, there exist consecutive read requests, some of which possibly access the same keys due to the zipfian distribution. For the consecutive sequence of read requests, only one write coordination and execution process is needed for all reads accessing the same key. Moreover, some keys are repeatedly written, leading to combined write coordination and write execution among Cassandra nodes. Thus, the last round has the largest throughput and its write/read cost ratio is different from the third round. Due to the large number of write requests and the effect of the background maintenance, the key distribution does not have much influence on the second round.

Next we consider the effect of consistent writes followed by immediate read requests. Different from the first group, the second group features more reads in the first two rounds. The write/read ratio and the requested key distribution are shown in Table \ref{tbl:morereads}. We can observe that the average latencies of reads (writes) stay close to each other in each round. The latency of writes at each round is significantly larger than that of reads. The total throughputs for all rounds are significantly smaller than that in Table \ref{tbl:morewrites}. This fact implies again the necessity of write batch processing. However, when the write consistency is very important, e.g., as in financial applications, write consistent will help.
\subsection{Discussion}%
\label{sec:discussion}%
As revealed by the experimental results, our latency estimation is not so accurate, though it is sufficient to validate dConssandra. That is, decomposing replication process into small steps and approximating by simple functions are feasible mechanisms of dConssandra. Exploiting better latency estimation techniques will improve the results. Furthermore, NoSQL stores generally support multi-tenancy, which implies an even more complicated load status, thus increasing the difficulty of latency estimation.

The immediate read/write, as we call it in the above sections, is not as immediate as indicated by the name. The minimum latency needed to process a read/write depends on the actual configuration of the system and the realtime workload of the node.

It is natural to consider that making consistent write and immediate read requests will lead to high throughput, if the workload consists is mostly reads with very few writes. However, as indicated by the latter two rows of Table \ref{tbl:morewrites} and the first two rows of Table \ref{tbl:morereads}, this is not correct. Making immediate write and consistent read actually lead to a better total throughput. Thus, further study is needed on how to decide the best bound choices for read/write requests based on the workload.\vspace{-10pt}
\section{Related Work}%
\label{sec:related}%
In distributed system with replication, replica consistency represents how replicas of the same logical data object correspond with each other, and with the ideal replica state when there is only one replica \cite{distributedsys}. In traditional database system, the consistency of ACID properties emphasizes the database complying with legal protocols defined by users under atomic transactions \cite{gray:consistency}. Work on consistency rationing \cite{rationing} tries and presents flexible models for suiting different cloud applications with operations under eventual consistency or ACID transaction guarantees at different moments. But eventual consistency and ACID transaction guarantees in fact concern different granules of operation processing.

Replica consistency is in a trade-off relation with availability and network partition tolerance \cite{brewer:cap}. For example, Dynamo \cite{dynamo} preserves availability and provides eventual consistency; while GFS \cite{gfs} underlying BigTable \cite{bigtable} trades off availability for strict consistency and provides one-copy equivalence consistency. In a system not encountering network partition, replica consistency can be degraded for lower response latency. For example, PNUTS \cite{pnuts} offers operations with version specifications. But version is not an easy parameter to specify, especially in a large-scale sharing scenario where multiple applications access the same data set. Another related work that studies the replication and latency problem controls latency by way of replica locality \cite{kadambi:where}. Although some NoSQL store like Cassandra \cite{cassandraPaper} provides multiple consistency choices, it actually guarantees strong consistency on the ALL mode and eventual consistency on the other modes through the read-repair mechanism.

There has been abundant research \cite{kadambi:where,keeton:disasters} on asynchronous replication exploring the trade-offs between replication frequency, application RPO (recovery point objective) demands, financial outlay by application owners, and possibly even multi-site replication. We are the first to explore the replica consistency versus latency trade-off for operation latency bounding. %\cite{tradeoffExp}

Paxos is a fault-tolerant protocol that enables strong consistency guarantee. Paxos execution incurs a long time, and thus recent work is devoted to finding a feasible implementation that provides high performance for data stores  \cite{paxosRSM, ibmPaxos}. Though performance is improved for Paxos implementation, the extra time needed to bring replicas consistent is inevitable as compared to eventual consistency. Some \cite{cops} borrows consistency concepts from the shared-memory architecture, e.g. causal consistency which requires a global view of system-wide operation sequences.

YCSB \cite{ycsb} is a cloud testing framework, which actually provides test for NoSQL stores like HBase, PNUTS, and Cassandra. YCSB++ \cite{ycsbplus} is an extended version that adds multi-test coordination, eventual consistency measurement, and multi-phase workloads to enhance performance understanding.%\vspace{-6pt}
\section{Conclusion}%
\label{sec:conclude}%
In this paper, we study the problem of providing the best consistency based on a user-specified latency bound. We present the dConssandra approach for dynamically bounding the operation response latency by trading off replica consistency in the NoSQL store. In dConssandra's staging and stepwise approach, the replication strategy is decomposed into stages so that latency can be controlled by stage recombination. The replication stages are then further decomposed into minute steps such that latency can be estimated and approximated by summation of simple linear functions. The proposed dCON algorithm maximizes the replica consistency while searching for the right step subset to recombine. dConssandra adopts a new storage architecture to support this staging and stepwise approach. Durability and failure tolerance are guaranteed in the process.

dConssandra is just a first step. We see several directions for future work. One direction is to improve the latency estimation for the operation execution process in a running NoSQL store. Techniques can be borrowed from the DB benchmarking and performance measurement area to improve the accuracy of the NoSQL operation latency estimation. Another direction for future work is to find a feasible way of integrating this replica consistency variance with the traditional transaction framework. Traditional transaction definition requires strong replica consistency, but we could also have consistency probability or other consistency statuses. From the application point of view, the future work also exists in exploring the possibility of charging by response latency in a NoSQL storage service. At a higher level, dConssandra's approach opens up a new dimension in providing storage service with regard to response latency and consistency. Exploring this new dimension is yet another interesting future work.%\vspace{-6pt}
\section*{Acknowledgment}
This work was supported in part by China national key project Unstructured Data Management System (2010ZX01042-002-002).\vspace{-5pt}

\bibliographystyle{IEEEtran}

%{\footnotesize
\bibliography{IEEEabrv,ref}
%}

\balance

\end{document}